%% file: giachero.LTD15.arxiv.tex
\documentclass[final]{article}
\usepackage{graphicx}
\usepackage{rotating}
\usepackage{amssymb}
\usepackage{mathptmx}
\usepackage[numbers]{natbib}
\usepackage{subscript}
\usepackage{gensymb} 
\usepackage{fancyhdr}
\usepackage{lastpage}

\usepackage[vcentering,dvips,left=3.2cm,right=3.2cm,top=4cm,bottom=4cm]{geometry}

\usepackage{authblk}

\makeatletter

\bibpunct{}{}{,}{s}{}{,}

\begin{document}

\newcommand{\hdblarrow}{H\makebox[0.9ex][l]{$\downdownarrows$}-}
\title{Critical Temperature tuning of Ti/TiN multilayer films suitable for low temperature detectors}


\author[1]{A.Giachero\footnote{tel.: +39 02-6448-2456, e-mail: Andrea.Giachero@mib.infn.it}}
\author[2]{P. Day}
\author[3]{P. Falferi}
\author[1]{M. Faverzani}
\author[1]{E. Ferri}
\author[4]{C. Giordano}
\author[4]{B. Marghesin}
\author[4]{F. Mattedi}
\author[5]{R. Mezzena}
\author[1]{R. Nizzolo}
\author[1]{A. Nucciotti}
\affil[1]{Universit\`a di Milano-Bicocca and INFN Milano-Bicocca, Italy}
\affil[2]{Jet Propulsion Laboratory, Pasadena, CA, U.S.A}
\affil[3]{Istituto di Fotonica e Nanotecnologie, CNR-Fondazione Bruno Kessler, Trento, Italy}
\affil[4]{Fondazione Bruno Kessler, Trento, Italy}
\affil[5]{Dipartimento di Fisica, Universit\`{a}  di Trento, Trento, Italy}


\date{\today}

\maketitle

\input{giachero.LTD15.text.tex}

\end{document}

%% file: giachero.LTD15.text.tex
\begin{abstract}
We present our current progress on the design and test of Ti/TiN Multilayer for use in Kinetic Inductance Detectors (KIDs). Sensors based on sub-stoichiometric TiN film are commonly used in several applications. However, it is difficult to control the targeted critical temperature $T_C$, to maintain precise control of the nitrogen incorporation process and to obtain a production uniformity. To avoid these problems we investigated multilayer Ti/TiN films that show a high uniformity coupled with high quality factor, kinetic inductance and inertness of TiN. These features are ideal to realize superconductive microresonator detectors for astronomical instruments application but also for the field of neutrino physics. Using pure Ti and stoichiometric TiN, we developed and tested different multilayer configuration, in term of number of Ti/TiN layers and in term of different interlayer thicknesses. The target was to reach a critical temperature $T_C$ around $(1\div 1.5)$~K in order to have a low energy gap and slower recombination time (i.e. low generation-recombination noise).  The results prove that the superconductive transition can be tuned in the $(0.5\div 4.6)$~K temperature range by properly choosing the Ti thickness in the $(0\div 15)$~nm range, and the TiN thickness in the $(5\div 100)$~nm range. 
\end{abstract}

\section{Introduction}
Kinetic Inductance Detectors (KIDs), proposed for the first time in 2003\cite{Day}, have developed rapidly over the last decade, and have by now reached performances that make them competitive with respect to other more mature technologies. Thanks to the opportunity to tune their geometry and to choose the material that is most suitable for the specific requirements of the experiment, they have been studied for applications in many astronomical instruments for radiation detection from sub-mm\cite{MUSIC} to gamma-ray\cite{Mazin,Cecil}. Recently they have been proposed also for particle detectors applicable to rare-events search such as the direct detection of dark matter\cite{Moore}, the neutrinoless double beta decay, the direct measurement of the neutrino mass\cite{Faverzani1,Faverzani2} and the measurement of the coherent neutrino-nucleus scattering.

Titanium Nitride (TiN), has been recently investigated as superconducting material, and has shown very good performances for KIDs. TiN resonators show, in particular, very high quality factors and very good optical coupling. These properties are mainly due to the large fraction of kinetic inductance of this material but also to its very low loss and high resistivity in the normal state\cite{Zmuidzinas,Vissers}. Furthermore, the amount of Nitrogen added to the Titanium film allows one to tune the transition temperature in the range $(0\div 4.5)$~K. The availability of superconducting thin films with critical temperature tuned for specific applications is a common issue for many low temperature detectors. Our target is to use films of Titanium Nitride to develop athermal detectors for the electron-capture decay endpoint measurement of the neutrino mass using Holmium (\textsuperscript{163}Ho) as source material\cite{Faverzani1,Faverzani2}. In order to have a low energy gap and slower recombination time (i.e. low generation-recombination noise) the films must have a critical temperature below $(1\div 1.5)$~K. We report in this work our first tests done at fabricating TiN films with $T_{c}<4.5K$ and we present the preliminary results obtained on resonators made using these films.

\section{Films production}
\begin{figure}[!t]
  \centering
  \includegraphics[width=0.8\textwidth]{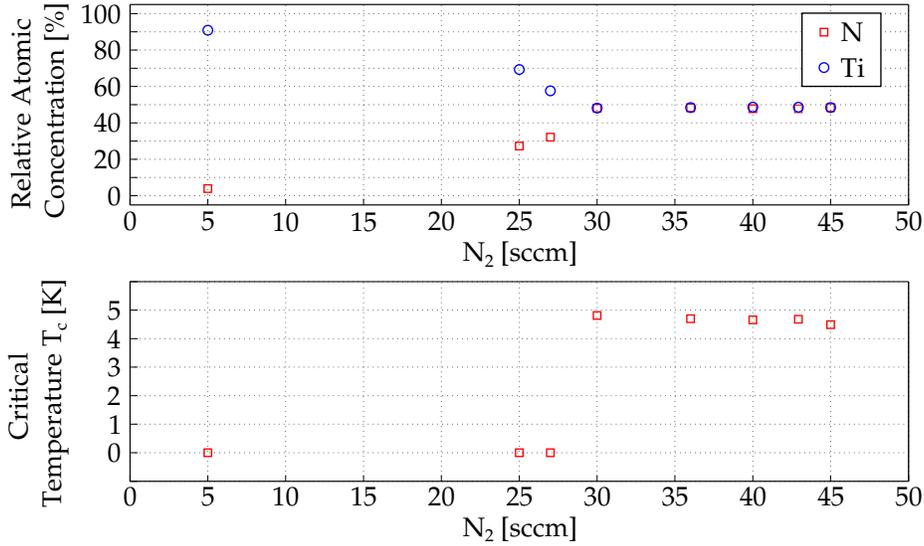}
  \caption{Top: Titanium and Nitrogen relative atomic concentration of the TiN\textsubscript{x} samples as a function of the N\textsubscript{2} flow rate. The atomic concentration ratios only slightly deviate from the stoichiometric value, except for N\textsubscript{2} flow rate of 27, 25 and 5 sccm, where values of respectively 0.56, 0.39 and 0.04 are reached. Bottom: the critical temperatures of TiN\textsubscript{x} samples as a function of the N\textsubscript{2} flow rate. The values are close to 4.5~K for films produced with a N\textsubscript{2} flow rate between 45 and 30 sccm. No transitions are observed below 30 sccm, $T_c=0$ in the plot. This could be an indication of how sharp is the transition from the high $T_c$ regime to the low one.}
  \label{fig:TiNConcentration}
\end{figure}

Our first attempt was to produce sub-stoichiometric TiN\textsubscript{x} films with our MRC Eclipse reactive sputtering system in order to investigate the dependence of the superconducting transition temperature on the $x$ parameter~\cite{Leduc}. The films were produced in the fabrication facilities of Fondazione Bruno Kessler (FBK). They were deposited in a mixture of Ar-N\textsubscript{2} gas from a pure Ti target. Within the Nitrogen concentration range explored (N\textsubscript{2} flow rate between 45~sccm and 5~sccm\footnote{Standard Cubic Centimeters per Minute, a Flow measurement term.}), the critical temperatures $T_C$ of the films were either close to $4.5~$K or were not observed at all. Auger Electron Spectroscopy (AES) were performed in FBK thus providing us detailed information on the elemental composition of the materials and on the chemical states of the surface atoms. All in-depth measurements showed a nearly stoichiometric composition of the TiN\textsubscript{x} in the range $(30\div 45)$~sccm (figure \ref{fig:TiNConcentration}). The samples produced with N\textsubscript{2} flow rate of 27, 25 and 5~sccm showed no transition. The chemical composition of the 27 and 25~sccm samples were in fact very similar to that of Ti\textsubscript{2}N, which is not a superconductor, whereas it is very close to pure Titanium for the sample with 5~sccm. These results are not really unexpected because of two important considerations: the change in the critical temperature occurs over a very narrow range of Nitrogen concentration, but on the other hand it turned out to be very difficult to control the Nitrogen concentration in the film by changing the N\textsubscript{2} flow rate parameter in the deposition recipe. This difficulty in controlling $T_C$ is a great challenge. A possible solution is to consider a different approach using a multilayer of pure Ti and stoichiometric TiN.


The idea is to fabricate films constituted by a sequence of pure Titanium and stoichiometric TiN layers. We based our choice on the well known proximity effect in bilayers\cite{Silvert}. Many experimental publications\cite{Werthamer,Yu} show that the $T_{c}$ of a superconducting material can be reduced by the superposition of a normal metal, or a metal which exhibits a superconductive transition at much lower temperature. In this case some Cooper pairs leak into the normal metal, thus reducing the pair density in the superconducting metal and its $T_{c}$. The magnitude of this deficiency depends on the thicknesses of the two layers: by adjusting these thicknesses one is able to tune the $T_{c}$ between two limiting values, i.e. the transition temperatures of the single layers. Titanium and TiN have a $T_{c}$ of about 0.4~K and 5~K, respectively thus, by adjusting the relative thicknesses of these layers, it is possible to tune the $T_{c}$ between these two values.

Our multilayer have been produced by the superposition of a number (max 8) of bilayers Ti/TiN with Ti at the bottom interface and TiN always on the top. The films have been sputtered on high resistivity Silicon wafers which have been first cleaned and etched with hydrofluoric acid (HF) in order to remove the native oxide. The films were deposited at a temperature of 400\degree~and the thin thicknesses of Ti layers have been obtained by changing the power and the deposition time in the recipe. The wafers were finally diced into slices of width from 0.5~mm to 1.5~mm and length of around 1~cm, then the slices were mounted on a copper support and electrically connected. 

\begin{figure}[!t]
  \centering
  \includegraphics[width=0.9\textwidth]{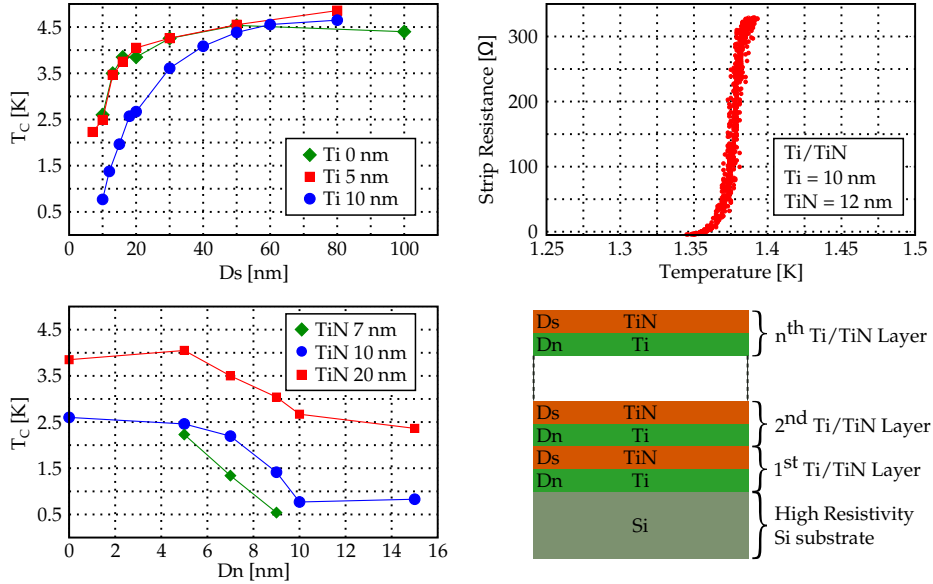}
  \caption{The left plots show the measured $T_{c}$ of the Ti/TiN multilayers as function of TiN (top-left) and Ti (bottom-left) single layers thicknesses. In the bottom-right the schematic of multilayers is also reported. The transitions are sharp, with a width of $\simeq$30~mK, as shown in the top-right figure, where data from multilayer Ti/TiN~=~10nm/12nm are illustrated. A good reproducibility and uniformity in the films was obtained.}
  \label{fig:TinTiNPlot}
\end{figure}

The DC 4-wires measurements of the electrical resistance were performed at the Cryogenics Laboratory of the University of Trento. The samples were installed in a dilution refrigerator (Janis JDry-100-ASTRA) with  base temperature of 17~mK. Figure \ref{fig:TinTiNPlot} shows the critical temperature as a function of the layer thicknesses. Results prove that the superconductive transition can be tuned in the $(0.5\div 4.6)$~K temperature range by properly choosing the Ti thickness in the $(0\div 15)$~nm range, and the TiN thickness in the $(7\div 100)$~nm range. We observed that the $T_{c}$ can be lowered by reducing the TiN thickness once we fixed the thickness of Ti. It is also possible to keep fixed the TiN layers and to increase the thickness of Ti. In this case we observed that if on one hand the $T_{c}$ is more sensitive to the composition of the bilayers in the $(5\div 10)$~nm range, on the other hand it is nearly stable for Ti thickness $<5 nm$ and $>10 nm$. In order to check uniformity, we measured $T_c$ across the wafer on different slices cut from the edge and from the center and found that the critical temperature varied by no more than 1\%. In the case of TiN\textsubscript{x} we found variation up to 20\%. These results are compatible with the measurements recently presented by Vissers \textit{et al.}~\cite{Vissers2}.

\section{RF Characterization}
We designed multilayer Ti/TiN detector array chips of 16 individual detectors with frequencies distributed in the range from 4 to 5.5~GHz and with a variety of coupling quality factors. The resonator geometry (lumped-element resonator) consists of two interdigital capacitors (IDC) connected with a coplanar strip (CPS) transmission line that works as inductor\cite{Faverzani1}. Each resonator is capacitively coupled on one side to a coplanar waveguide (CPW) line that is used for the readout.

\begin{figure}[!t]
  \centering
  \includegraphics[width=0.9\textwidth]{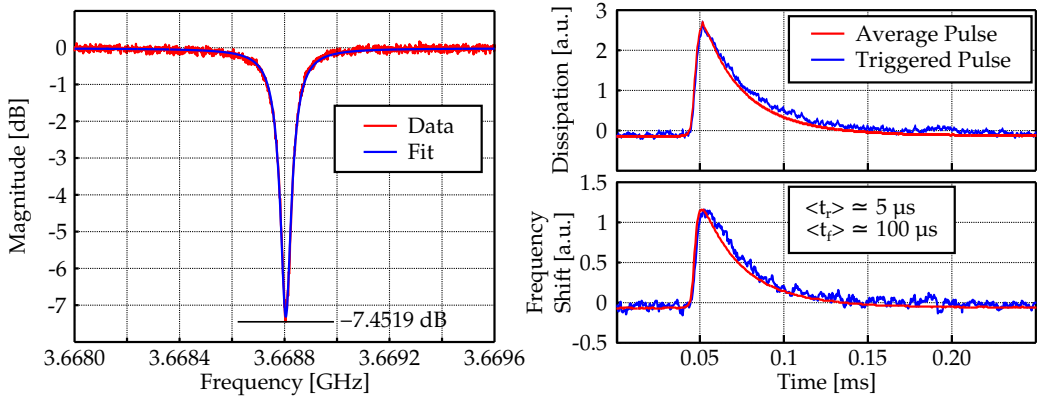}
  \caption{The left plot shows an example of resonance frequency measurement, $f_{res}=3.6688$~GHz. By fitting the resonances to Lorentzian lineshapes, the internal quality factor TiN film was found around $Q_i=10^5$. The right plot shows the observed pulse response for the produced Ti/TiN multilayers from an aluminum x-ray fluorescence source. Data are digitized at 2.5~MS/s and projected into the frequency and dissipation quadrature.
}
  \label{fig:RFChar}
\end{figure}

The array was mounted in a dilution fridge (Oxford MX 40, testbed at the Cryogenics Laboratory of the University of Milano-Bicocca) and cooled down to temperature as low as 100~mK. For each resonator, we measured the resonance frequency ($f_{res}$) by using a Vector Network Analyzer (HP 8753E, 30~kHz~$\div$~6~GHz) remotely controlled. In figure \ref{fig:RFChar} (left) is shown an example of one resonance with its fit reported. The kinetic inductance fraction $\alpha$ of our resonators was estimated comparing simulations on the designed layout with the experimental results, as well as by fitting the shift of the resonance frequency as a function of the temperature with the theoretical prevision~\cite{Gao}, as well as by comparing the resonance frequencies of TiN resonators with that of our aluminum resonators having the same geometry. Both TiN and multilayer Ti/TiN films showed a typical value 0.5 for $\alpha$, which correspond to an increase by more than one order of magnitude, compared with the Al version.

In order to test the detector’s response to the release of energy, the devices were illuminated with an aluminum x-ray fluorescence source on the top of the detectors. A probe signal, generated by a microwave synthesizer, was used to excite the resonators. The transmitted signal was amplified with a cryogenic HEMT amplifier, mounted on the 4K~stage and, subsequently, with a room-temperature amplifier. Homodyne mixing with the original excitation signal is accomplished employing an IQ-mixer. The resulting I and Q signals were acquired by a commercial data acquisition board with a sample rate $f_s=2.5$~MHz, and projected offline into the frequency and dissipation quadrature signals. Pulses due to X-ray absorption events were easily identified in the detector output timestream, as it is shown in figure \ref{fig:RFChar} (right) for a multilayer composed by 8 layers of Ti/TiN (176~nm of thickness, Ti/TiN=10~nm/12~nm). The data analysis of the acquired pulses is currently in progress and it will be treated in a separate publication.

\section{Conclusions}
We produced and tested sub-stoichiometric TiN and Ti/TiN multilayer films suitable for a next generation of experiments in the neutrino physics field. In the case of sub-stoichiometric TiN films it is difficult to control the target $T_c$ and, additionally, they show large non-uniformity. In the case of Ti/TiN multilayer films we observed that the target $T_c$ can be tuned by reducing the TiN thickness once we fixed the thickness of Ti or keeping fixed the TiN layers and increasing the thickness of Ti. Results prove that the superconductive transition can be tuned in the $(0.5\div 4.6)$~K temperature range by choosing properly the Ti thickness in the $(0\div 15)$~nm range and the TiN thickness in the $(5\div 100)$~nm range with a good reproducibility and uniformity.

\section*{Acknowledgments}
This work is supported by Fondazione Cariplo through the project \textit{Development of Microresonator Detectors for Neutrino Physics} (grant \textit{International Recruitment Call 2010}, ref. 2010-2351).

%% file: giachero.LTD15.arxiv.bbl
\begin{thebibliography}{99}

\bibitem{Day}
P. K. Day \textit{et al.}, {\it Nature} \textbf{425}, 817-821 (2003). 

\bibitem{MUSIC}
P. R. Maloney \textit{et al.}, {\it Proc. SPIE} \textbf{7741}, 77410F (2010). 

\bibitem{Mazin}
B.A. Mazin \textit{et al.}, {\it Appl. Phys. Lett.}, \textbf{89}, 222507 (2006). 

\bibitem{Cecil}
T. Cecil \textit{et al.}, {\it Appl. Phys. Lett.}, \textbf{101}, 032601 (2012). 

\bibitem{Zmuidzinas}
J. Zmuidzinas, {\it Annual Review of Condensed Matter Physics}, \textbf{3}, 169-214 (2012). 

\bibitem{Moore}
J. Gao, \textit{et al.}, {\it Appl. Phys. Lett.}, \textbf{101}, 142602 (2012). 

\bibitem{Gao}
J. Gao, \textit{et al.}, {\it NIMA}, \textbf{559}, 585-587 (2006). 

\bibitem{Leduc}
H. G. Leduc \textit{et al.}, {\it Appl. Phys. Lett.}, \textbf{97}, 102509 (2012); 

\bibitem{Vissers}
M. R. Vissers, \textit{et al.}, {\it Appl. Phys. Lett.}, \textbf{97}, 232509 (2010). 

\bibitem{Vissers2}
M. R. Vissers, \textit{et al.}, {\it Appl. Phys. Lett.}, \textbf{102}, 232603 (2013). 

\bibitem{Faverzani1}
M. Faverzani, \textit{et al.}, {\it J. Low. Temp. Phys.} \textbf{167}, 1041–1047 (2012). 

\bibitem{Faverzani2}
M. Faverzani, \textit{et al.}, {\it NIMA}, \textbf{167}, 492–494 (2013). 

\bibitem{Silvert}
W. Silvert, {\it Journal of Low Temperature Physics} \textbf{20}, 439, (1975)

\bibitem{Werthamer}
N. R. Werthamer, {\it Physical Review} \textbf{132}, 2440, (1963).

\bibitem{Yu}
L. Yu, N. Newman, and J. M. Rowell, {\it IEEE Transactions on Applied Superconductivity} \textbf{12}, 1795, (2002). 


\end{thebibliography}
